\documentclass{pasj00}
\draft

\begin{document}
\SetRunningHead{M. Sato and H. Asada}
{Transiting Extrasolar Planet and Companion in Eccentric Orbit} 
\Received{}
\Accepted{}

\title{Transiting Extrasolar Planet with a Companion: \\
Effects of Orbital Eccentricity and Inclination} 

\author{Masanao \textsc{Sato}, 
and  
Hideki \textsc{Asada}} %
\affil{Faculty of Science and Technology, Hirosaki University, 
Hirosaki, Aomori 036-8561}


%


\maketitle

\begin{abstract}
Continuing work initiated in an earlier publication 
[Sato and Asada, PASJ, 61, L29 (2009)], 
we consider light curves influenced by 
the orbital 
inclination and 
eccentricity of a companion in orbit around 
a transiting extrasolar planet 
(in a planet-satellite system or a 
hypothetical 
true binary). 
We show that the semimajor axis, eccentricity 
and inclination angle 
of a 
`moon' 
orbit around the host planet 
can be determined by transit method alone. 
For this purpose, 
we present a formulation for the parameter determinations 
in a small-eccentricity approximation 
as well as in the exact form. 
As a result, the semimajor axis is expressed 
in terms of observables such as brightness changes, 
transit durations 
and intervals in light curves. 
We discuss also a narrow region of parameters that 
produce a mutual transit by an extrasolar satellite. 
\end{abstract}

\section{Introduction}
It is of general interest to 
discover a second Earth-Moon system. 
Detections of such an extrasolar planet with a satellite 
(or 
hypothetical 
binary planet systems that do not exist in the Solar System) 
and probing the nature of such objects 
will bring important information to 
planet (and satellite) formation theory 
(e.g., 
Williams et al. 1997, 
Jewitt and Sheppard 2005, 
Canup and Ward 2006, 
Jewitt and Haghighipour 2007). 
If a giant planet with a (perhaps Earth-size) rocky satellite 
were located at a certain distance from their host star, 
the satellite may be habitable and show vegetation, 
though these issues are out of the scope of this paper. 

It is not clear whether 
the IAU definition for planets in the Solar System
can be applied to extrasolar planets as it is. 
The IAU definition in 2006 is as follows:
A planet is a celestial body that 
(a) is in orbit around the Sun, 
(b) has sufficient mass so that it assumes a hydrostatic equilibrium 
(nearly round) shape, and 
(c) has cleared the neighborhood around its orbit.

Regarding (c), the Earth can be called a planet, mostly because 
the common center of mass (COM) of the Earth-moon system is 
below the surface of the Earth. 
On the other hand, the COM of the Pluto-Charon system is located 
above the surfaces of these objects. 
Therefore, it is 
interesting 
to determine 
the COM position of a planet-companion system. 
In order to determine it, 
we have to know the true (not apparent) distance 
between the two objects. 
For this reason, Sato and Asada (2009) considered 
extrasolar mutual transits, 
as a complementary method of measuring not only 
the radii of two transiting objects but also 
their separation (See Sato and Asada 2009 also on 
detection probabilities of extrasolar mutual transits 
and a possible limit by Kepler Mission). 
As a particular case, 
a short separation binary, which has a rapidly orbiting companion, 
gives us a unique opportunity to measure the true separation of the binary, 
whereas a long separation one gives us only the apparent separation. 
Their work is very limited, however,  in the sense that they assume 
circular orbits 
and also coplanar orbits as $I = 90$ degrees 
for both planet and moon. 
Clearly it is important to take account of the orbital 
inclination and 
eccentricity.  
The main purpose of this paper is to study effects of 
the orbital 
inclination and 
eccentricity of a companion 
on extrasolar mutual transits. 

Since the first detection of a transiting extrasolar planet 
(Charbonneau et al. 2000), photometric techniques have been 
successful (e.g., Deming et al. 2005 
for probing atmosphere, 
Ohta et al. 2005, Winn et al. 2005, Gaudi \& Winn 2007, 
Narita et al. 2007, 2008 for measuring spin-orbit alignment angle). 
In addition to 
{\it COROT}\footnote{http://www.esa.int/SPECIALS/COROT/}, 
{\it Kepler}\footnote{http://kepler.nasa.gov/} 
is monitoring about $10^5$ stars with 
expected 
20 ppm 
($=2 \times 10^{-5}$) 
photometric differential sensitivity 
for stars of V=12. 
This will marginally enable the detection of a moon-size object. 
In fact, {\it COROT} detected a transiting super-Earth 
(Leger et al. 2009, Queloz et al. 2009). 

Sartoretti and Schneider (1999) first 
suggested a photometric detection of extrasolar satellites. 
Cabrera and Schneider (2007) developed a method 
based on the imaging of a planet-companion as an unresolved 
system (but resolved from its host star) 
by using planet-companion mutual transits and mutual shadows. 
As an alternative method, 
timing offsets for a single eclipse have been investigated 
for eclipsing binary stars 
as a perturbation of transiting planets around the center of mass 
in the presence of the third body 
(Deeg et al. 1998, 2000, Doyle et al. 2000). 
It has been recently extended toward detecting `exomoons'  
(Szab{\'o} et al. 2006, 
Simon et al. 2007, 
Kipping 2009a, 2009b). 
Sato and Asada (2009) investigated 
effects of mutual transits by an extrasolar planet with a companion 
on light curves.  
In particular, they studied 
how the effects depend on the 
companion's orbital 
velocity. 
Furthermore, extrasolar mutual transits were discussed 
as a complementary method of measuring the system's parameters  
such as a planet-companion's separation 
and thereby of identifying them as a true binary, 
planet-satellite system or others.

Their method has analogies in classical ones for eclipsing binaries 
(e.g., Binnendijk 1960, Aitken 1964). 
A major difference is that 
occultation of one faint object by the other 
transiting a parent star causes an apparent {\it increase}  
in light curves, 
whereas eclipsing binaries make a decrease. 
What is more important is that, in both cases where one faint object 
transits the other and vice versa, 
changes are made in the light curves due to mutual transits 
even if no light emissions come from the faint objects. 
In a single transit, on the other hand, thermal emissions from 
a transiting object at lower temperature make a difference 
in light curves during the secondary eclipse, when the object 
moves behind a parent star as observed for instance for 
HD 209458b 
(Deming et al. 2005).  

This paper is organized as follows. 
In section 2, we consider effects of the orbital 
inclination and 
eccentricity of a companion on light curves. 
For simplicity, 
henceforth, such a companion 
orbiting around a host planet is called a `moon' 
even if it is not a satellite but a component 
of a hypothetical binary planet. 
In section 3, we present a formulation 
for parameter determinations. 
Some numerical examples are also presented. 
Section 4 is devoted to the conclusion.

\section{Effects of the Orbital Eccentricity of 
a Transiting `Moon' on Light Curves} 
\subsection{Approximations and notation}
The time duration of a transit, say 
a few hours, 
is much longer than the orbital period of an extrasolar planet, 
say a few days or greater. 
In one transit, the effect of the motion of the moon 
is much larger than that of the planet orbiting around a star.  
During the transit, therefore, we employ a constant velocity 
approximation only for the orbital motion of a planet-moon system 
around their host star. 
For a short separation case, on the other hand, 
we take account of the eccentric orbit of the moon, 
because the orbital period of such a moon around the planet may be 
comparable to (or shorter than) the timescale of the transit. 

The co-planar assumption that the orbital plane of a moon 
around its primary object is the same as that of the planet 
in orbit around the host star seems reasonable 
because it seems that planets are born from fragmentations 
of a single proto-stellar disk 
and thus their spins and orbital angular momentum are 
nearly parallel to the spin axis of the disk. 
Irregular satellites such as Triton, however,  
have significant inclinations
presumably through capture processes. 
This requires that we should include the effect of orbital
inclinations.  
We assume only the moon's orbital inclination, 
because it makes substantial effects 
on mutual transit light curves. 
Inclinations of the planet's orbital plane have been well understood 
and already observed (Charbonneau et al. 2000).

Here we list our assumptions for clarity. 
\begin{itemize}
\item The inclination angle of the COM of the planet and moon 
is fixed at 90 degrees. 
\item We take account of the inclination angle of the moon's orbit.  
\item We assume that the planet-moon COM has a constant velocity 
(during a transit by the planet in front of the host star). 
\item The longitude of ascending node of the moon equals zero. 
\item The planet-moon COM orbit has zero eccentricity. 
\item We assume no limb darkening effects. 
\end{itemize}

For an eccentric orbit, we generally have the Kepler equation as 
\begin{equation}
t=t_0 + \frac{T}{2\pi} (u-e \sin{u}) , 
\label{kepler}
\end{equation}
where $t_0$, $T$, $e$ and $u$ denote the time of periastron passage, 
orbital period, eccentricity and eccentric anomaly, respectively 
(e.g., Danby 1988, Roy 1988, Murray and Dermott 1999).  
In the following, we use the true anomaly $f$ instead of 
the eccentric anomaly $u$ 
(e.g., Danby 1988, Roy 1988, Murray and Dermott 1999). 
They are related by 
\begin{equation}
\tan\frac{f}{2} = \sqrt{\frac{1+e}{1-e}} \tan\frac{u}{2} . 
\end{equation}
The distance between the orbiting body and a focus of the ellipse 
is written as 
\begin{equation}
r = \frac{a(1-e^2)}{1+e\cos f} . 
\end{equation}

We use these equations for describing a moon orbiting 
around a planet. 
We denote the 
mean motion 
of the moon in orbit around the primary 
as $n_m \equiv 2\pi/T_m$, 
where the subscript $m$ means the moon's quantity. 
The subscript $p$ denotes the planet's quantity. 

\begin{figure}
    \FigureFile(140mm,130mm){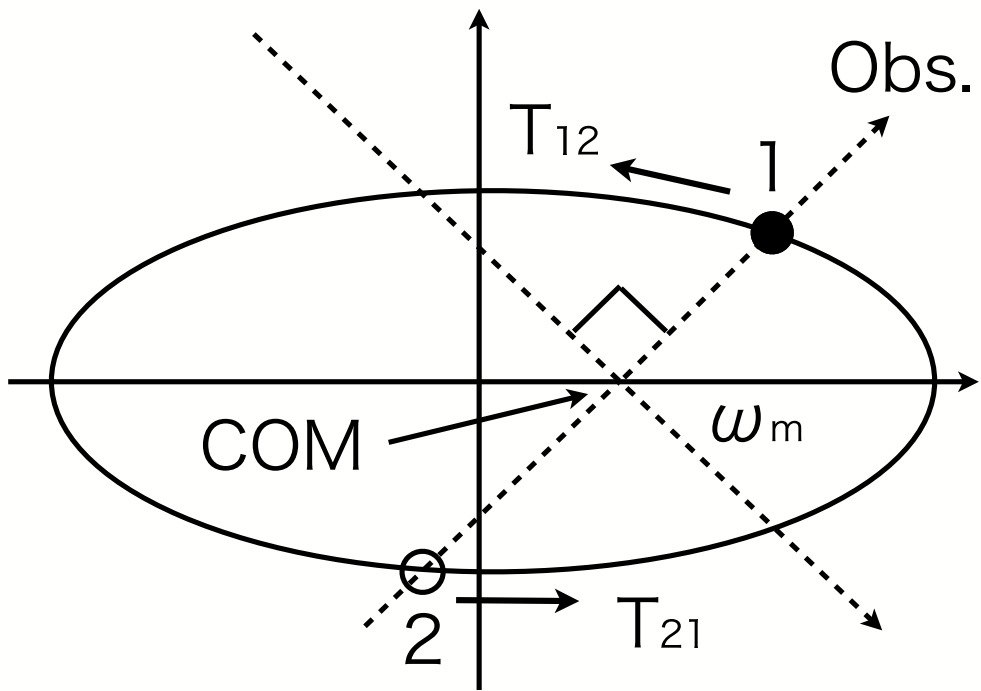}
\caption{
Direction of the line of sight. It is denoted as $\omega_m$, 
which is the argument of pericenter. 
}
\label{config}
\end{figure}

For investigating transits, we need 
the transverse position $x$ and velocity $v$ 
of each object. 
We denote those of 
the COM for planet-moon systems  
as $x_{CM}$ and $v_{CM}$, respectively, 
where the origin of $x$ is chosen as the center of the star.  
We assume $v_{CM}$ as constant during the transit. 
The position and velocity of each planet 
with mass $M_p$ and $M_m$  
in the planet-moon system as 
$x_i$ and $v_i$ ($i=p, m$), respectively. 
The direction of the observer's line of sight 
is specified by the argument of pericenter  
as an angle denoted by $\omega_m$ (See also Fig. $\ref{config}$). 
We express the transverse position as 
\begin{eqnarray}
x_{CM}
&=&
v_{CM}  (t-t_{CM}) ,  \\
\label{xCM}
x_p 
&=& 
x_{CM}+a_p [ (\cos u - e_m) \cos\omega_m + \sqrt{1-e_m^2} \sin u \sin\omega_m ] , 
\label{x1-u}
\\
x_m 
&=& 
x_{CM}-a_m [ (\cos u - e_m) \cos\omega_m + \sqrt{1-e_m^2} \sin u \sin\omega_m ] ,
\label{x2-u}
\end{eqnarray}
where 
the semimajor axis of the orbit 
of each object around their COM 
is denoted by $a_i$, 
and 
$t_{CM}$ means the time when the binary's common center of mass 
passes in front of the center of the host star. 
In terms of the true anomaly, they are rewritten as 
\begin{eqnarray}
x_p 
&=& 
x_{CM}+a_p \frac{(1-e_m^2) \cos(f+\omega_m)}{1 + e_m \cos f} , 
\label{x1-f}
\\
x_m 
&=& 
x_{CM}-a_m \frac{(1-e_m^2) \cos(f+\omega_m)}{1 + e_m \cos f} . 
\label{x2-f}
\end{eqnarray}
(See Table 1 for a list of parameters and their definition). 

The azimuthal velocity of the secondary object around the primary 
is 
\begin{eqnarray}
V_f &=& r \frac{df}{dt}  
\nonumber\\
&=& 
a_{pm}n_m 
\frac{1+e_m\cos f}{\sqrt{1-e_m^2}} , 
\label{Vf}
\end{eqnarray}
where $f$ denotes the true anomaly (Murray and Dermott 2000). 
Here, 
$a_{pm}$ and $e_m$ denote the semimajor axis and eccentricity 
of the moon's orbit with respect to the host planet. 
We assume that both the eccentricity of COM orbit vanishes 
and the inclination angle of COM equals 90 degrees. 
Hence we can avoid a careful treatment of 
``sky-projected transverse position''.

\subsection{Transits in light curves} 
We denote the intrinsic stellar luminosity as $L$. 
The apparent luminosity $L^{'}$ due to mutual transits 
is expressed as 
\begin{equation}
L^{'} = L \times \frac{S-\Delta S}{S} , 
\label{L}
\end{equation}
where 
$S=\pi R_s^2$, 
$S_p=\pi R_p^2$,
$S_m=\pi R_m^2$,
$\Delta S=S_p+S_m-S_{pm}$. 
Here, $R_s$, $R_p$ and $R_m$ denote the radii of 
the host star, planet and moon, 
and 
$S_{pm}$ denotes the area of 
the apparent overlap between them, 
which is seen from the observer. 
Without loss of generality, we assume that 
the primary is larger than the secondary 
as $R_p \geq R_m$.

\subsection{Effects on light curves}
We investigate light curves 
by mutual transits due to planet-moon systems.  
The orbital velocity is of the order of $a_{pm}n_m$. 
Therefore, we have two cases; 
$v_{CM} < a_{pm}n_m$ and $v_{CM} > a_{pm}n_m$. 

\begin{figure}
    \FigureFile(140mm,130mm){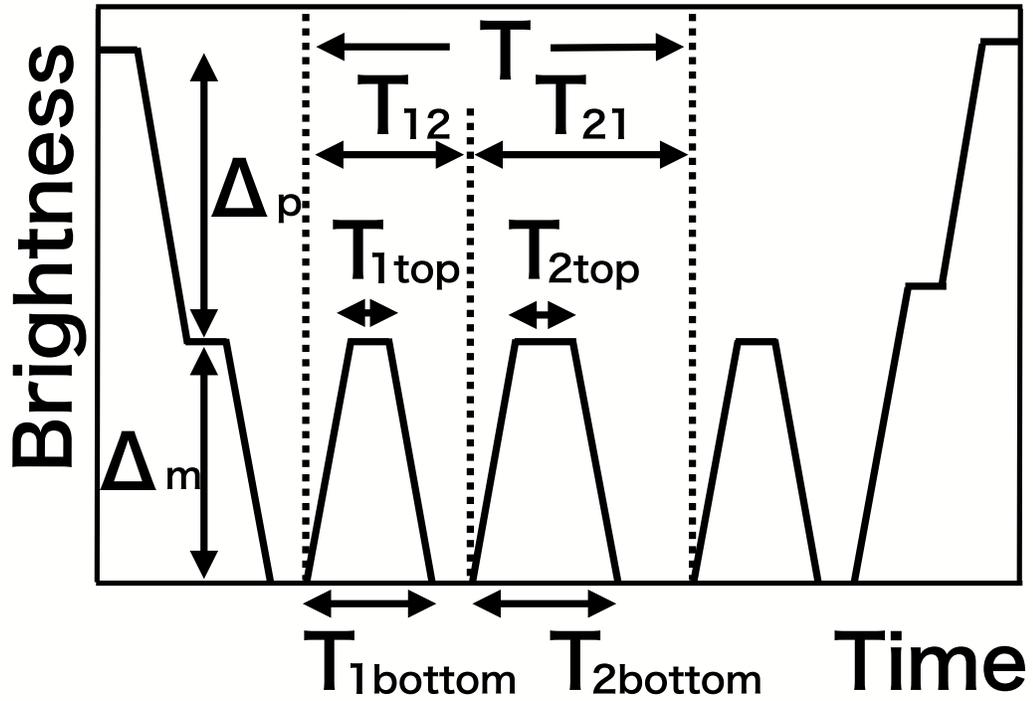}
\caption{
Schematic figure of a light curve due to 
a mutually transiting planet and moon 
in front of their host star. 
}
\label{schematic}
\end{figure}

The dimensionless ratio of the moon's orbital velocity 
to the planet's one is defined as 
\begin{equation}
W \equiv \frac{a_{pm}n_m}{v_{CM}} . 
\label{W}
\end{equation}
If $v_{CM} < a_{pm} n_m$, 
we call it a fast case. 
If $v_{CM} > a_{pm} n_m$, we call it a slow one.  
The Earth-Moon ($W=0.03$), Jupiter-Ganymede ($W=0.8$) 
and Jupiter-Io ($W=1.3$) systems represent 
slow, marginal and fast cases, respectively. 
Figure $\ref{schematic}$ shows a schematic light curve 
by mutual transits. 

\begin{figure}
    \FigureFile(120mm,120mm){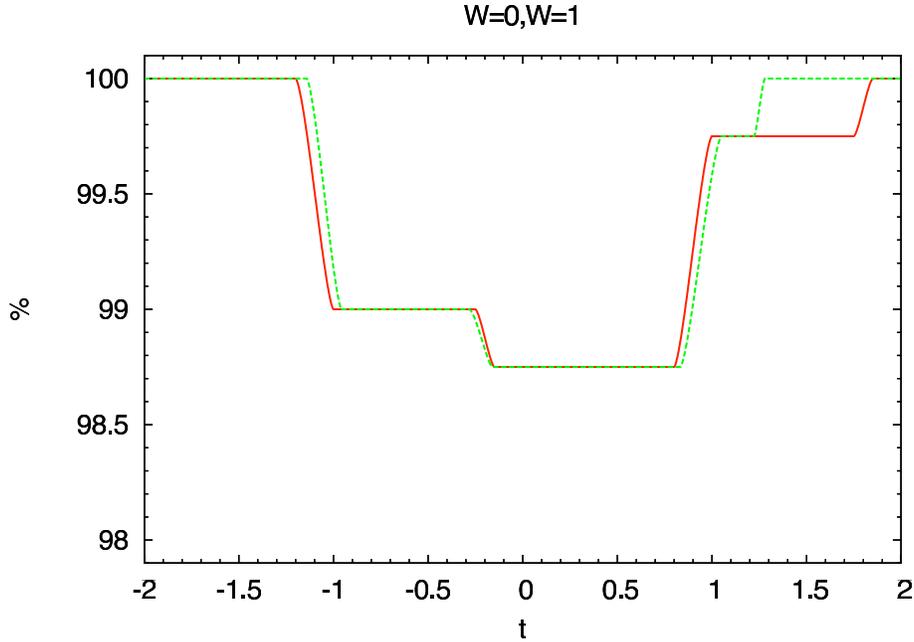}
\caption{
Light curves: 
Solid red one denotes 
the zero limit of a `moon' orbital motion as a reference 
($W \equiv a_{pm}n_m/v_{CM} = 0$). 
Dashed green one is a marginal spin case (large separation) for $W=1$ 
(Sato and Asada, 2009). 
The vertical axis denotes the apparent luminosity 
(in percents). 
The horizontal one is time in units of 
the half crossing time of the star by 
the COM of the binary, defined as $R_s/v_{CM}$. 
If one takes $M_s=M_{\odot}$, $R_s=R_{\odot}$ and $v_{CM}=30$ km/s 
(namely, 1 AU distance from the host star), 
$t=1$ corresponds to $\approx 6$ hours. 
For simplicity, we assume the binary with a common mass density, 
a radius ratio as 
$R_s: R_p: R_m =20: 2: 1$, 
and $a_{pm}/R_s=0.9$. 
}

\label{lightcurve-1}
\end{figure}

Fig. $\ref{lightcurve-1}$ shows a slow case in circular motion, 
where we assume $R_s: R_p: R_m = 20: 2: 1$. 
We assume also the same mass density for the two transiting objects 
and hence obtain $a_p: a_m = 1: 8$. 

\begin{figure}
    \FigureFile(120mm,90mm){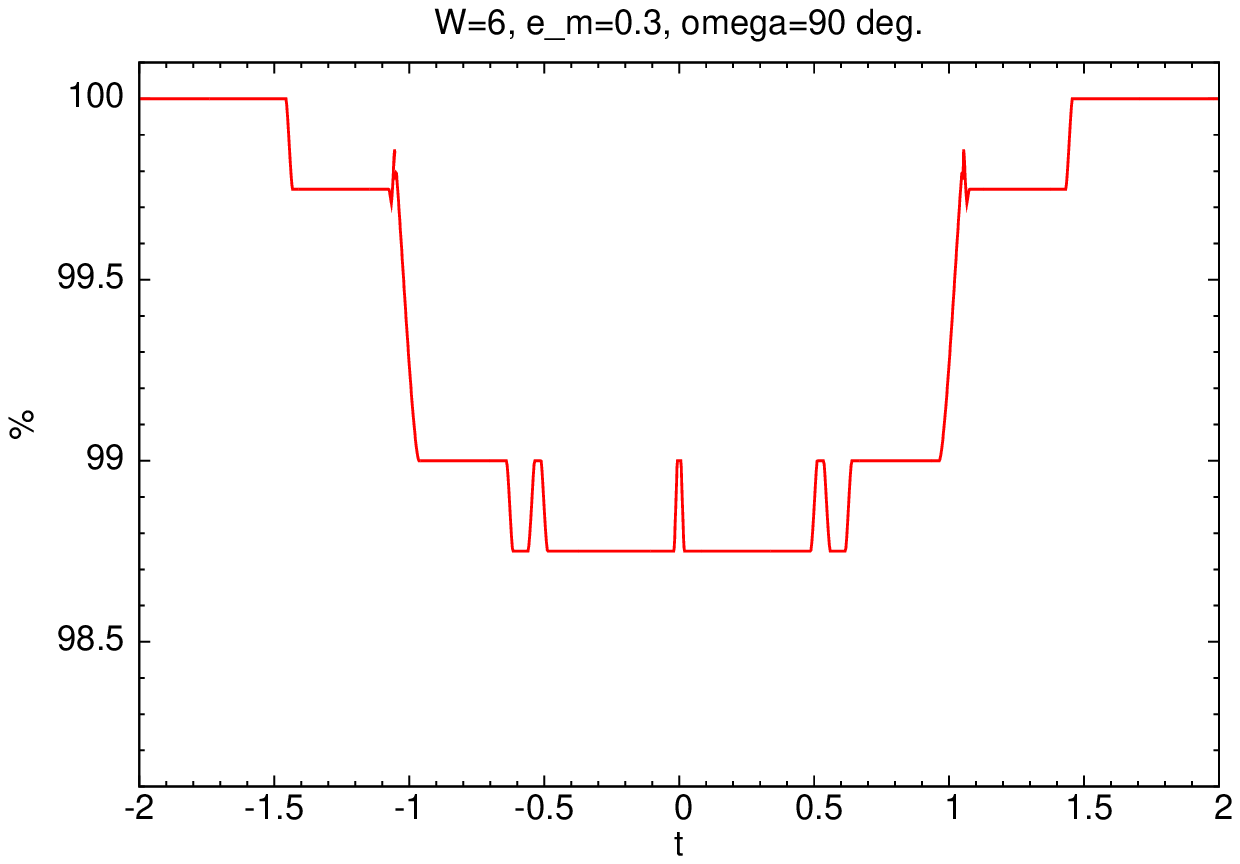}
\\
    \FigureFile(120mm,90mm){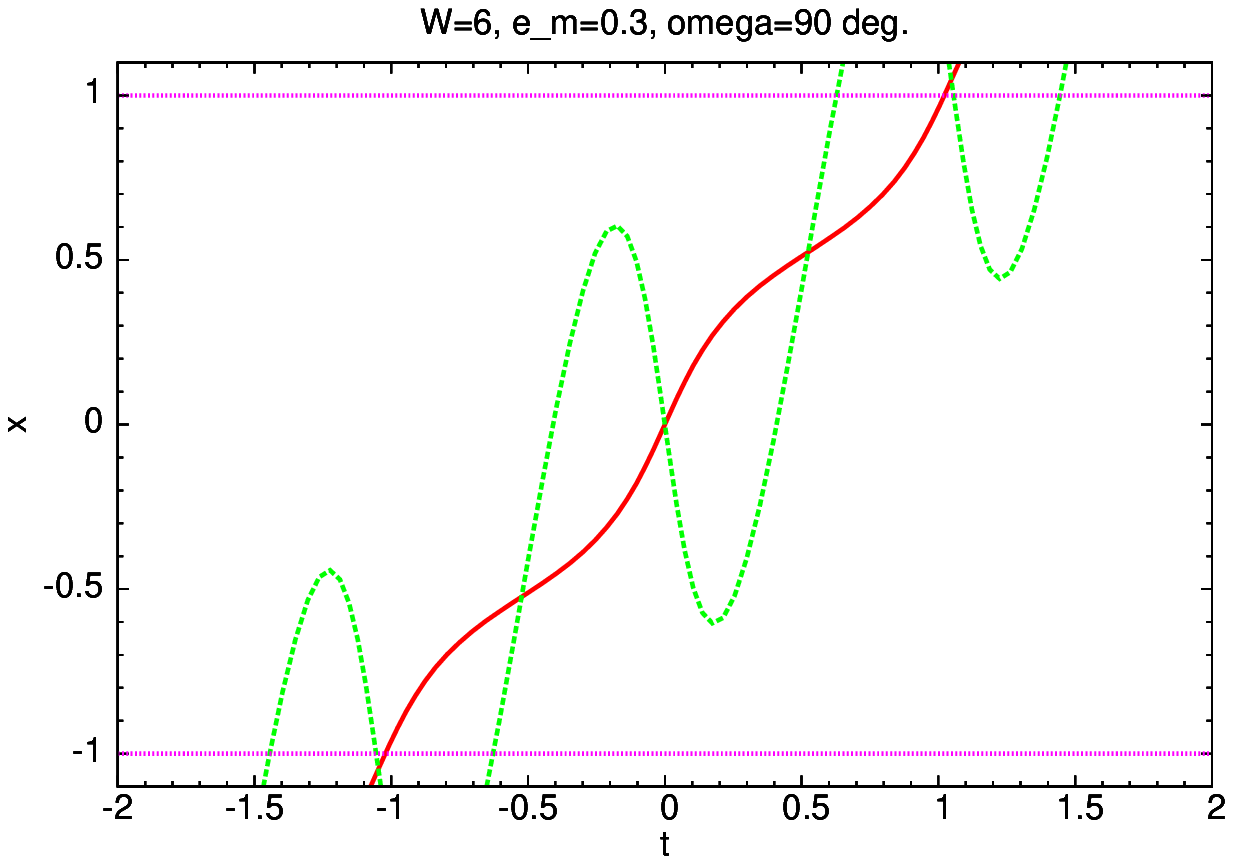}
\caption{
Top panel: a light curve for a fast case (small separation) 
as $W=6$ and $e_m=0.3$. 
The radius and separation are the same as those 
in Fig. $\ref{lightcurve-1}$. 
The observer's direction is $\omega_m=\pi/2$. 
Brightness fluctuations appear with 
$T_{1top}/T_m=0.013$, $T_{1bottom}/T_m=0.022$, 
$T_{2top}/T_m=0.039$, $T_{2bottom}/T_m=0.067$, 
$T_{12}/T_m=0.51$ and $T_{21}/T_m=0.49$ (normalized by $T_m$, 
the orbital period of the moon: $T_m=T_{12}+T_{21}$). 
We obtain $T_r \sim 1.7$ from Eq. ($\ref{ecosPsi}$) 
and thus $e_m\sin\omega_m \sim 0.3$, 
whereas Eq. ($\ref{deltaT2}$) approximately gives us $e_m\cos\omega_m \sim 0$. 
Therefore, we recover well 
the parameters as $e_m \sim 0.3$ and $\omega_m \sim \pi/2$, 
even in the linear approximation in $e_m$. 
Finally, Eq. ($\ref{a1bottom2}$) tells $a_{pm}/R_s \sim 0.9$. 
Bottom panel: motion of each body 
in the direction of $x$ normalized by $R_s$ 
(solid red for the primary and dotted green for the secondary). 
When one faint object transits or occults the other 
in front of the host star, 
mutual transits occur 
and a `hill' appears in the light curve. 
}
\label{lightcurve-2}
\end{figure}

\begin{figure}
    \FigureFile(120mm,90mm){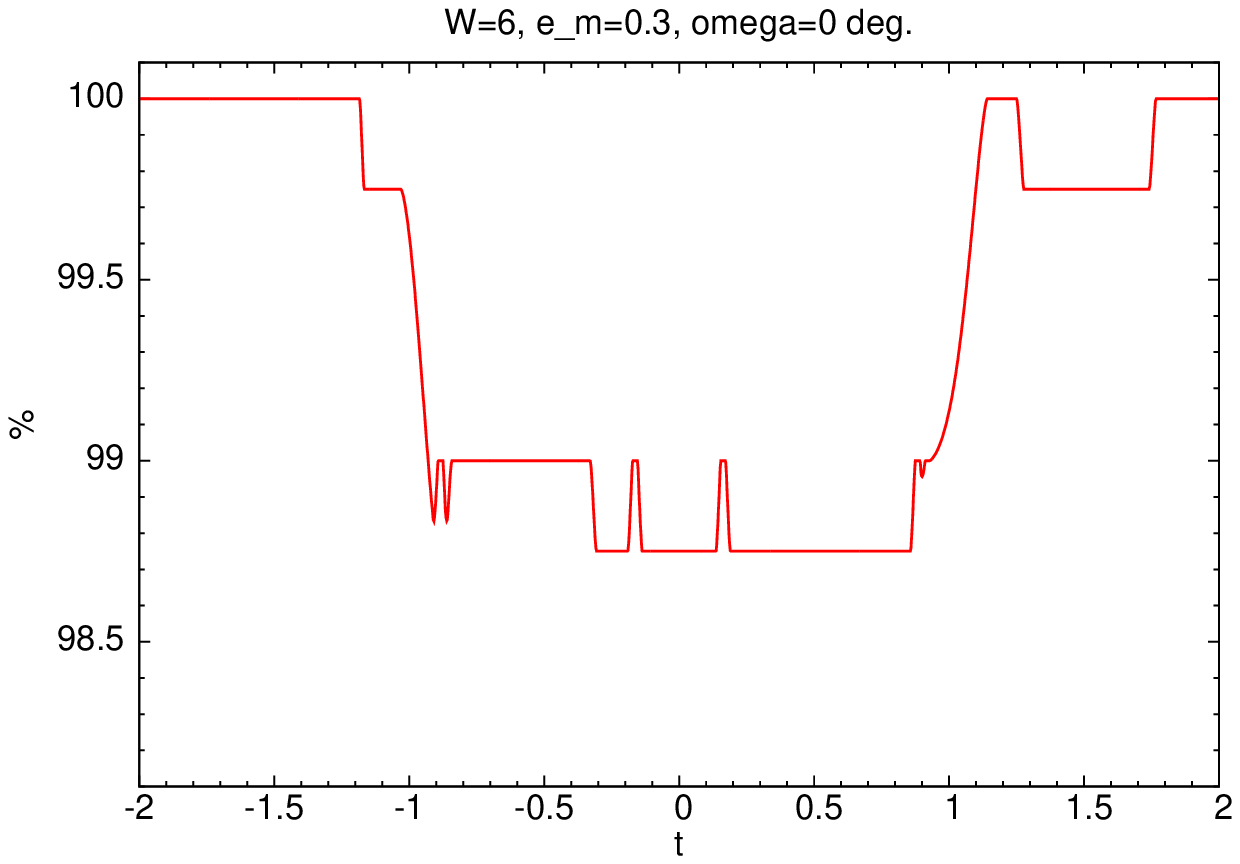}
\\
    \FigureFile(120mm,90mm){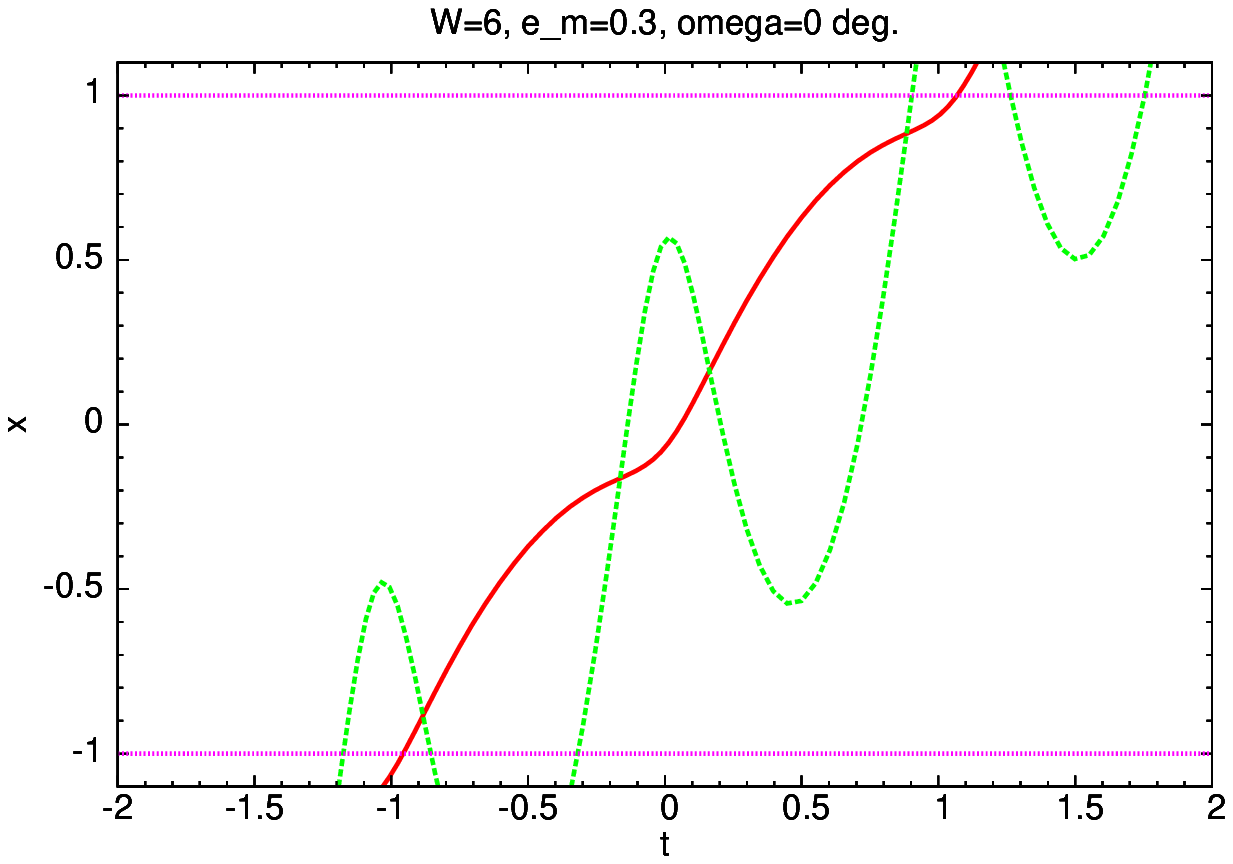}
\caption{
Top panel: a light curve for all the parameters same as 
those in Fig. $\ref{lightcurve-2}$, except for  
the observer's direction as $\omega_m=0$. 
Brightness fluctuations appear with 
$T_{1top}/T_m=0.018$, $T_{1bottom}/T_m=0.018$, 
$T_{2top}/T_m=0.050$, $T_{2bottom}/T_m=0.050$, 
$T_{12}/T_m=0.69$ and $T_{21}/T_m=0.31$.  
We obtain $T_r \sim 1.0$ from Eq. ($\ref{ecosPsi}$) 
and thus $e_m\sin\omega_m \sim 0$, 
whereas Eq. ($\ref{deltaT2}$) approximately gives us $e_m\cos\omega_m \sim 0.3$. 
Therefore, we recover well the parameters 
as $e_m \sim 0.3$ and $\omega_m \sim 0$, 
even in the linear approximation in $e_m$. 
Finally, Eq. ($\ref{a1bottom2}$) tells $a_{pm}/R_s \sim 0.9$. 
Bottom panel: motion of each body 
(solid red for the primary and dotted green for the secondary). 
}
\label{lightcurve-3}
\end{figure}

\begin{figure}
    \FigureFile(120mm,90mm){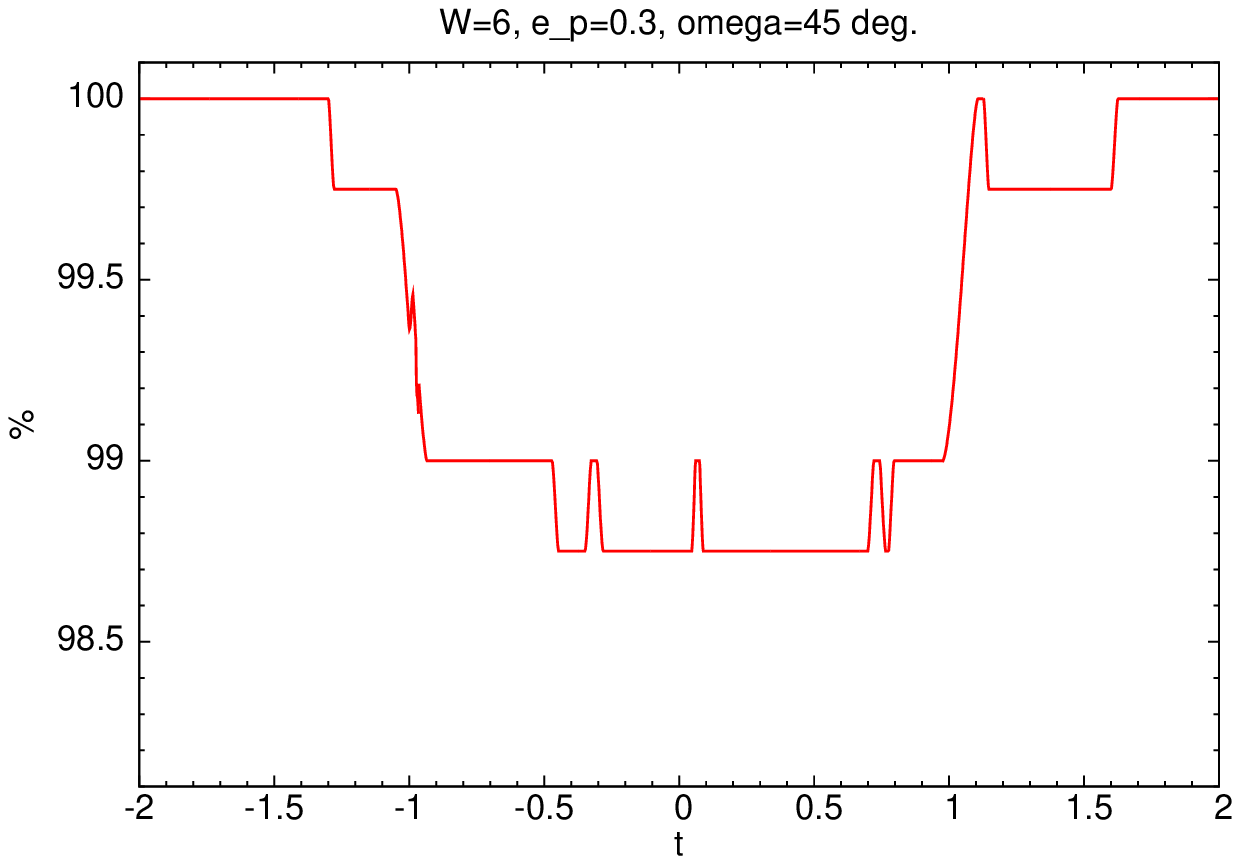}
\\
    \FigureFile(120mm,90mm){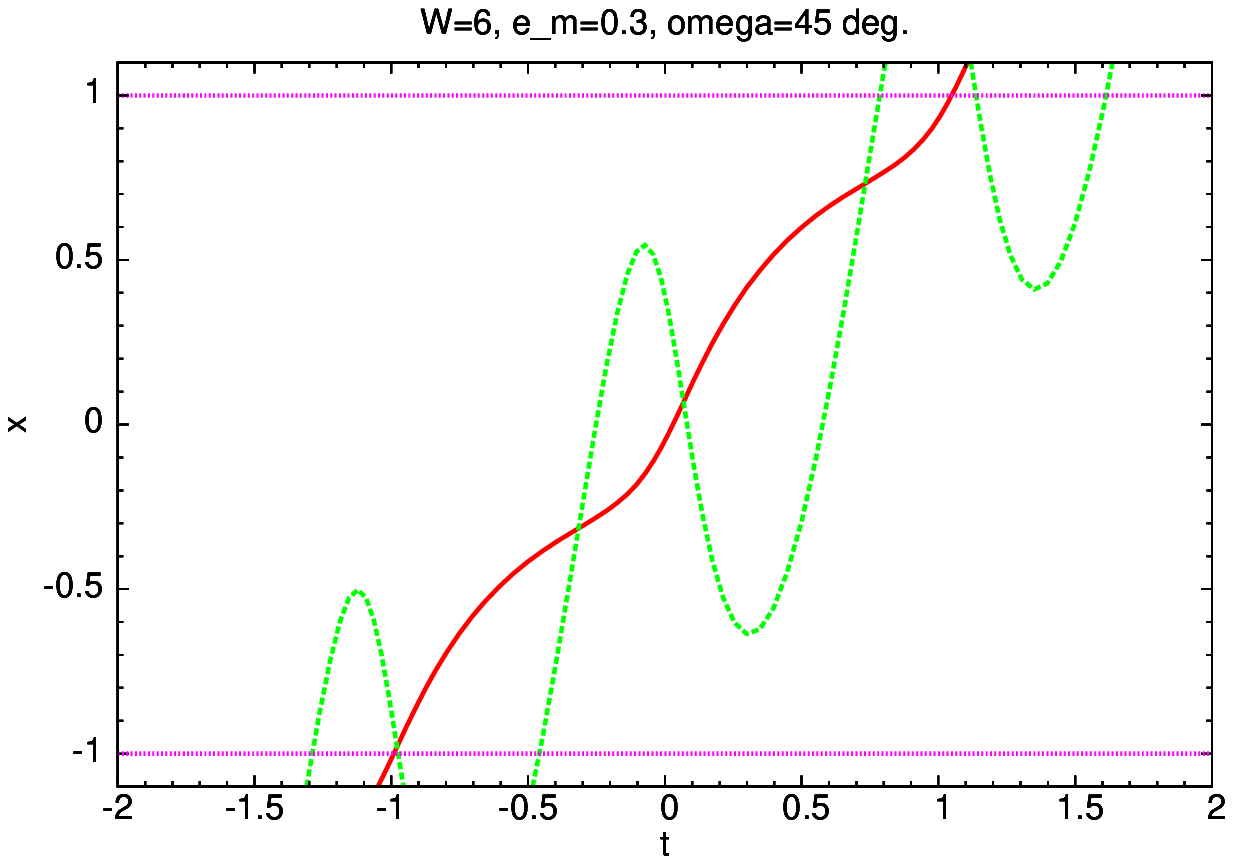}
\caption{
Top panel: a light curve for all the parameters same as 
those in Fig. $\ref{lightcurve-2}$, except for 
the observer's direction as $\omega_m=\pi/4$. 
Brightness fluctuations appear with  
$T_{1top}/T_m=0.015$, $T_{1bottom}/T_m=0.022$, 
$T_{2top}/T_m=0.041$, $T_{2bottom}/T_m=0.062$, 
$T_{12}/T_m=0.62$ and $T_{21}/T_m=0.38$.  
We obtain $T_r \sim 1.5$ from Eq. ($\ref{ecosPsi}$) 
and thus $e_m\sin\omega_m \sim 0.2$, 
whereas Eq. ($\ref{deltaT2}$) approximately gives us $e_m\cos\omega_m \sim 0.2$. 
Therefore, we recover well the parameters 
as $e_m \sim 0.3$ and $\omega_m \sim \pi/4$, 
even in the linear approximation in $e_m$. 
Finally, Eq. ($\ref{a1bottom2}$) tells $a_{pm}/R_s \sim 0.9$. 
Bottom panel: motion of each body 
(solid red for the primary and dotted green for the secondary). 
}
\label{lightcurve-4}
\end{figure}

Eccentric orbit cases ($W=6$ and $e_m=0.3$) are shown by 
Figs. $\ref{lightcurve-2}$, $\ref{lightcurve-3}$ 
and $\ref{lightcurve-4}$ 
($\omega_m=\pi/2$, $0$ and $\pi/4$, respectively).  
Some parameters are chosen so that effects in the figures 
can be distinguished by eye, 
though such an event is unlikely to be detected 
by current observations as discussed later. 
For generating the ingress and egress of the various parts 
of the lightcurve, we do not use a linear interpolation but 
compute numerically the apparent overlap area between the objects. 
Here we assume the same configuration 
except for the observer's line of sight.   
For simplicity, we take $t_0 = t_{CM} = 0$ in these figures.  

These figures show also the transverse positions of 
transiting objects with time, 
which would help us to understand the chronological changes 
in the light curves. 
In particular, 
it can be understood that such characteristic patterns appear 
only when two objects are in front of the star and 
one of them transits (or occults) the other.

\section{Formulation for Parameter Determinations by Transit Method}
\subsection{Parameter determinations from transit observations alone} 
In all the above cases, the amount of decrease in light curves 
or the magnitude of fluctuations gives the ratios among the radii of 
the star and two faint objects $(R_s, R_p, R_m)$.  
The decrease ratios in the apparent brightness 
due to transits by the planet and moon
are written as 
\begin{eqnarray}
\Delta_p &=& \left(\frac{R_p}{R_s}\right)^2 , 
\label{Delta1}
\\
\Delta_m &=& \left(\frac{R_m}{R_s}\right)^2 . 
\label{Delta2}
\end{eqnarray}
The stellar radius $R_s$ (and mass $M_S$) 
are known for instance by its spectral type.
Hence, the radii are expressed in terms of observables 
$R_s$, $\Delta_p$ and $\Delta_m$ as 
\begin{eqnarray}
R_p &=& R_s \sqrt{\Delta_p} , 
\label{r1}
\\
R_m &=& R_s \sqrt{\Delta_m} . 
\label{r2}
\end{eqnarray} 
We define the ratio between the brightness changes by the two objects
as 
\begin{equation}
\Delta \equiv \frac{\Delta_m}{\Delta_p} .
\end{equation}

\noindent
{\bf Circular Orbit and Orbital Inclination: }\\
First, we discuss a circular orbit  
in order to simply explain our idea. 
For more rigorous treatment of eccentric orbits, 
please see below, where we will finally give expressions for 
determining the separation $a_{pm}$. 

Behaviors of apparent light curves depend on $W$. 
Therefore, 
$a_{pm}n_m$ (as its ratio to $v_{CM}$) can be obtained 
(Sato and Asada 2009). 
Seager and Mall{\'e}n-Ornelas (2003) presents 
an analytic solution of parameter determinations 
for a single transit in the circular orbit case. 
Their solution can be used for our case 
of mutual transits by a planet and moon 
in front of a host star. 
In our case, their equations are rewritten as follows. 
The duration of the `flat part' of the transit ($t_{Fm}$) 
is described by 
\begin{equation}
\sin\left( \frac{t_{Fm} \pi}{T_m} \right) 
= \frac{1}{a_{pm} \sin I_m} \sqrt{(R_p-R_m)^2 - (a_{pm} \cos I_m)^2} , 
\end{equation}
whereas the total transit duration ($t_{Tm}$) is done by 
\begin{equation}
\sin\left( \frac{t_{Tm} \pi}{T_m} \right) 
= \frac{1}{a_{pm} \sin I_m} \sqrt{(R_p+R_m)^2 - (a_{pm} \cos I_m)^2} , 
\label{sintT}
\end{equation}
where $I_m$ denotes the orbital inclination angle of the moon. 
By combining these equations, the impact parameter ($b_{pm}$) 
can be derived as 
\begin{eqnarray}
b_{pm} &\equiv&
\frac{a_{pm}}{R_p} \cos I_m 
\nonumber\\
&=&
\sqrt{\frac{(1-\sqrt{\Delta})^2 
- [\sin^2(t_{Fm}\pi/T_m)/\sin^2(t_{Tm}\pi/T_m)] 
(1+\sqrt{\Delta})^2}{1-[\sin^2(t_{Fm}\pi/T_m)/\sin^2(t_{Tm}\pi/T_m)]}} .
\end{eqnarray}
The ratio $a_{pm}/R_p$ can be derived directly from 
Eq. ($\ref{sintT}$) as 
\begin{equation}
\frac{a_{pm}}{R_p} = 
\sqrt{\frac{(1+\sqrt{\Delta})^2 
- b_{pm}^2 [1-\sin^2(t_{Tm}\pi/T_m)]}
{\sin^2(t_{Tm}\pi/T_m)}} .
\end{equation}
Therefore, one can obtain $a_{pm}$ as 
$R_s \times (R_p/R_s) \times (a_{pm}/R_p)$.

With $a_{pm}$ and $n_m$ in hand, 
one can thus estimate the total mass of the binary 
by $GM_{tot}=n_m^2 a_{pm}^3$ from Kepler's third law, 
where $G$ denotes the gravitational constant. 
The orbital velocity $a_{pm}n_m$ gives the mutual force 
between the binary.

If we assume also that the mass density is common 
for two objects constituting the binary 
(this may be reasonable especially for similar size objects 
as $R_p \sim R_m$), 
each mass is determined as $M_p = R_p^3 (R_p^3 + R_m^3)^{-1} M_{tot}$ 
and $M_m = R_m^3 (R_p^3 + R_m^3)^{-1} M_{tot}$, respectively. 
Therefore, the orbital radius of each body around the COM 
is obtained as $a_p = R_m^3 (R_p^3 + R_m^3)^{-1} a_{pm}$ 
and $a_m = R_p^3 (R_p^3 + R_m^3)^{-1} a_{pm}$, respectively. 
At this point, importantly, the two objects can be identified as 
a true binary ($a_p>R_p$) or planet-satellite system ($a_p<R_p$). 
However, a gaseous giant planet with a rocky satellite 
would exhibit largely different densities and this may 
be one of the most likely scenarios.

In a slow spin case, on the other hand, 
the apparent separation $a_{\perp}$ (normal to our line of sight) 
is determined as $a_{\perp} = T_{12} v_{CM}$ from measuring the time lag
$T_{12}$ between the first and second 
transits because $v_{CM}$ is known above (Sato and Asada 2009). 

\noindent
{\bf Eccentric Orbit and Edge-on Case:}\\
Henceforth, we take account of the orbital eccentricity 
of a moon for an edge-on case. 
In this case, intervals between neighboring ``hills'' 
are not constant because of the eccentricity. 
However, a time duration between three successive ``hills'' 
is nothing but the orbital period of the moon. 
Therefore, one can measure the period $T_m$. 
We obtain $n_m$ as 
\begin{equation}
n_m=\frac{2\pi}{T_m} . 
\label{omega}
\end{equation}

A key idea for determining the eccentricity is as follows. 
As for timescales, we have two observable ratios 
as $T_2/T_1$ and $T_{12}/T_{21}$. 
The former is the ratio between the widths of neighboring hills,  
whereas the latter is that between the transit intervals. 
On the other hand, we have two additional parameters 
$e_m$ and $\omega_m$ to be determined. 
Importantly, the number of measurable ratios is the same as 
that of the parameters that we wish to determine. 
In principle, therefore, the above two ratios may allow us to determine 
the two parameters $e_m$ and $\omega_m$, separately. 
This will be discussed in detail below. 

To be more precise, 
the full width of a ``hill'' at top and bottom 
are expressed as (See also Figure $\ref{schematic}$) 
\begin{eqnarray}
T_{top} &=& \frac{2(R_p-R_m)}{V_f}, 
\label{Ttop}
\\
T_{bottom} &=& \frac{2(R_p+R_m)}{V_f} . 
\label{Tbottom}
\end{eqnarray}
Only for symmetric binaries ($R_p=R_m$), 
we have $T_{top}=0$ and thus true spikes. 
Otherwise, truncated spikes (or ``hills'') appear. 

For the primary transit, where the moon moves 
in front of the planet, we have 
$f=\pi/2 - \omega_m$. 
{}From Eqs. ($\ref{Vf}$), ($\ref{Ttop}$) and ($\ref{Tbottom}$),  
therefore, we obtain 
\begin{eqnarray}
T_{1 top} &=& \frac{2(R_p-R_m)}{a_{pm}n_m} 
\frac{\sqrt{1-e_m^2}}{1+e_m\sin\omega_m} , 
\label{T1top}
\\
T_{1 bottom} &=& \frac{2(R_p+R_m)}{a_{pm}n_m} 
\frac{\sqrt{1-e_m^2}}{1+e_m\sin\omega_m} .  
\label{T1bottom}
\end{eqnarray}
For the circular orbit $e_m=0$, 
the second factors in the R.H.S. of these expressions 
become the unity and the first factors recover 
the case for $e_m=0$ (See Sato and Asada 2009). 

For the secondary transit, where the moon moves 
behind the planet, we have 
$f=3\pi/2 - \omega_m$. 
From Eqs. ($\ref{Vf}$), ($\ref{Ttop}$) and ($\ref{Tbottom}$),  
therefore, we obtain 
\begin{eqnarray}
T_{2 top} &=& \frac{2(R_p-R_m)}{a_{pm}n_m} 
\frac{\sqrt{1-e_m^2}}{1-e_m\sin\omega_m} , 
\label{T2top}
\\
T_{2 bottom} &=& \frac{2(R_p+R_m)}{a_{pm}n_m} 
\frac{\sqrt{1-e_m^2}}{1-e_m\sin\omega_m} .  
\label{T2bottom}
\end{eqnarray}

We immediately obtain from Eqs. ($\ref{T1top}$), 
($\ref{T1bottom}$), ($\ref{T2top}$) and ($\ref{T2bottom}$), 
\begin{eqnarray}
T_r &\equiv& 
\frac{T_{2 top}}{T_{1 top}} 
\nonumber\\
&=& 
\frac{T_{2 bottom}}{T_{1 bottom}}
\nonumber\\
&=&\frac{1+e_m\sin\omega_m}{1-e_m\sin\omega_m} .
\label{Tr}
\end{eqnarray}

Equation ($\ref{Tr}$) is rewritten as 
\begin{eqnarray}
e_m\sin\omega_m &=& \frac{T_r-1}{T_r+1} 
\nonumber\\
&\equiv& T_R, 
\label{ecosPsi}
\end{eqnarray}
where the R.H.S. can be determined by observations alone. 
Once either $e_m$ or $\omega_m$ is known, Eq. ($\ref{ecosPsi}$) 
determines the other. 

Next, we consider the time interval between the primary 
and secondary transits. 
In order to compute such an interval, one can use 
the Kepler's second law (the constant areal velocity). 
After lengthy but straightforward calculations, 
the area swept from the primary transit ($f=\pi/2 - \omega_m$) 
till the secondary ($f=3\pi/2 - \omega_m$) becomes 
\begin{eqnarray}
S_{12}&=& 
a_mb_m \times \left[ 
\frac{\pi}{2}+\arcsin H 
+\frac12 \sin(2\arcsin H) \right] , 
\label{S12}
\end{eqnarray}
where 
$b_m$ denotes the semiminor axis of the moon's elliptic orbit and 
$H$ is defined as 
\begin{equation}
H=e_m\sqrt{\frac{\cot^2\omega_m}{1-e_m^2+\cot^2\omega_m}} . 
\label{H}
\end{equation}

By using Eq. ($\ref{ecosPsi}$) in Eq. ($\ref{H}$) 
for eliminating $\cot\omega_m$, 
we obtain  
\begin{equation}
H=\sqrt{\frac{e_m^2-T_R^2}{1-T_R^2}} . 
\label{H2}
\end{equation}

It is convenient to use $T_{12}/T$ instead of $T_{12}/T_{21}$, 
because $T_{12}/T$ 
gives a simpler expression than 
$T_{12}/T_{21}$, which is used in the above explanation 
of the key idea. 
The total area is $S=\pi a_m b_m$. 
Therefore, we find 
\begin{eqnarray}
\frac{T_{12}}{T}&=&\frac{S_{12}}{S}
\nonumber\\
&=&\frac12 
+\frac{1}{\pi}\arcsin H 
+\frac{1}{\pi} H \sqrt{\frac{1-e_m^2}{1-T_R^2}} . 
\label{Tratio}
\end{eqnarray}
This includes $e_m$ without $\omega_m$. 
Hence by using this relation, one can determine separately 
the eccentricity by measuring the time intervals. 
We should note that Eq. ($\ref{Tratio}$) is valid 
even for a general case 
with a certain inclination angle.

Here, we consider a small eccentricity approximation, 
which may be useful for a quicker estimation of the parameters. 
For small $e_m$, Eq. ($\ref{H}$) is expanded as 
\begin{equation}
H=e_m \cos\omega_m+O(e_m^3) .  
\label{H-expansion}
\end{equation}
Substitution of this into Eq. ($\ref{Tratio}$) 
gives us 
\begin{eqnarray}
\frac{T_{12}}{T_m}
&=&\frac12 
+\frac{2e_m}{\pi}\cos\omega_m
+O(e_m^3) . 
\label{Tratio-expansion}
\end{eqnarray}
The correction to the circular case ($T_{12}/T_m=1/2$) 
is $2e_m \cos\omega_m/\pi \sim 0.7 e_m \cos\omega_m$. 
Even for $\omega_m\sim 0$ for instance, 
it leads to seven percents for $e_m=0.1$. 

We define the difference between $T_{12}$ and $T_{21}$ as 
\begin{equation}
\delta T \equiv T_{12}-T_{21} .  
\label{deltaT}
\end{equation}
Replacement $\omega_m$ by $\omega_m+\pi$ 
changes Eq. ($\ref{Tratio-expansion}$) into  
\begin{eqnarray}
\frac{T_{21}}{T_m}
&=&\frac12 
-\frac{2e_m}{\pi}\cos\omega_m
+O(e_m^3) . 
\label{Tratio-expansion2}
\end{eqnarray}

By using Eqs. ($\ref{Tratio-expansion}$) and 
($\ref{Tratio-expansion2}$), 
we obtain 
\begin{equation}
e_m \cos\omega_m = \frac{\pi \delta T}{4 T_m} + O(e^3) . 
\label{deltaT2}
\end{equation}

Let us consider every cases separately. 
  
(I) If and only if the L.H.S. of Eqs. ($\ref{ecosPsi}$) 
and ($\ref{deltaT2}$) vanish, $e_m$ does.  
This means a circular motion 
and thus the observer's direction $\omega_m$ becomes meaningless. 

(II) For a case when the L.H.S. of Eq. ($\ref{ecosPsi}$) vanishes 
but that of Eq. ($\ref{deltaT2}$) does not, 
we find $e_m \neq 0$ and $\sin\omega_m = 0$, namely 
\begin{equation}
\omega_m=0 \: (\mbox{mod}\: \pi). 
\end{equation}
Eq. ($\ref{deltaT2}$) immediately gives 
\begin{equation}
e_m=\frac{\pi \delta T}{4 T_m} .
\end{equation}
Hence, the orbital eccentricity is determined. 

(III) If the L.H.S. in Eq. ($\ref{deltaT2}$) vanishes 
but that in ($\ref{ecosPsi}$) does not, 
we find $e_m \neq 0$ and $\cos\omega_m = 0$, namely 
\begin{equation}
\omega_m=\frac{\pi}{2} \: (\mbox{mod}\: \pi). 
\end{equation}
Eq. ($\ref{ecosPsi}$) immediately gives 
\begin{equation}
e_m=T_R .
\end{equation}
Hence, the orbital eccentricity is measured. 

(VI) A general case in which 
the L.H.S. of neither Eqs. ($\ref{ecosPsi}$) nor ($\ref{deltaT2}$) 
vanish: 
By dividing Eq. ($\ref{deltaT2}$) by Eq. ($\ref{ecosPsi}$), 
we obtain 
\begin{equation}
\cot\omega_m = \frac{\pi}{4} \frac{\delta T}{T_m} \frac{1}{T_R} , 
\label{tanPsi}
\end{equation}
where the R.H.S. can be determined by observations alone 
and hence this equation gives us the observer's direction $\omega_m$. 
By substituting the determined $\omega_m$ into Eq. ($\ref{ecosPsi}$), 
one can find the value of the eccentricity. 

Up to this point, $e_m$ and $\omega_m$ both are determined. 
Eqs. $(\ref{T1top})$ and $(\ref{T1bottom})$ are rewritten as 
\begin{eqnarray}
a_{pm} &=& \frac{T_m (R_p-R_m)}{\pi T_{1top}}
\frac{\sqrt{1-e_m^2}}{1+e_m\sin\omega_m} 
\nonumber\\
&=& \frac{T_m (R_p-R_m) (T_{1top}+T_{2top}) \sqrt{1-e_m^2}} 
{2\pi T_{1top}T_{2top}} 
\nonumber\\
&=& \frac{(R_p-R_m)}{2\pi} 
\left(\frac{T_m}{T_{1top}}+\frac{T_m}{T_{2top}}\right) 
\sqrt{1-e_m^2} , 
\label{a1top}
\\
a_{pm} &=& \frac{T_m (R_p+R_m)}{\pi T_{1bottom}}
\frac{\sqrt{1-e_m^2}}{1+e_m\sin\omega_m}  
\nonumber\\
&=& \frac{T_m (R_p+R_m) (T_{1bottom}+T_{2bottom}) \sqrt{1-e_m^2}} 
{2\pi T_{1bottom}T_{2bottom}} 
\nonumber\\
&=& \frac{(R_p+R_m)}{2\pi} 
\left(\frac{T_m}{T_{1bottom}}+\frac{T_m}{T_{2bottom}}\right) 
\sqrt{1-e_m^2} , 
\label{a1bottom}
\end{eqnarray}
respectively, where we used Eq. ($\ref{ecosPsi}$)
If and only if $R_p=R_m$, we obtain 
$T_{1top} = T_{2top} = 0$ and 
Eq. ($\ref{a1top}$) thus becomes undetermined, 
whereas Eq. ($\ref{a1bottom}$) is still well-defined. 

When one wishes to consider the secondary transit instead of the first, 
one can use 
\begin{eqnarray}
a_{pm} &=& \frac{T_m (R_p-R_m)}{\pi T_{2top}}
\frac{\sqrt{1-e_m^2}}{1-e_m\sin\omega_m} , 
\label{a2top}
\\
a_{pm} &=& \frac{T_m (R_p+R_m)}{\pi T_{2bottom}}
\frac{\sqrt{1-e_m^2}}{1-e_m\sin\omega_m} . 
\label{a2bottom}
\end{eqnarray}
They are obtained by replacing $\omega_m$ with $\omega_m+\pi$ 
in Eqs. ($\ref{a1top}$) and ($\ref{a1bottom}$). 
By noting $T_{1top} (1+T_R) = T_{2bottom} (1-T_R)$ 
and $T_{2top} (1+T_R) = T_{2bottom} (1-T_R)$, 
one can show that Eqs. ($\ref{a2top}$) and ($\ref{a2bottom}$) 
agree with Eqs. ($\ref{a1top}$) and ($\ref{a1bottom}$), respectively. 
By using one of these expressions, we can thus measure  
the semimajor axis of the eccentric orbit. 
In terms of the decrease in apparent brightness, 
Eqs. ($\ref{a1top}$) and ($\ref{a1bottom}$) are written as 
\begin{eqnarray}
\frac{a_{pm}}{R_s} &=& 
\frac{(\sqrt{\Delta_p}-\sqrt{\Delta_m})}{2\pi} 
\left(\frac{T_m}{T_{1top}}+\frac{T_m}{T_{2top}}\right) 
\sqrt{1-e_m^2} ,
\label{a1top2}
\\
\frac{a_{pm}}{R_s} &=&
\frac{(\sqrt{\Delta_p}+\sqrt{\Delta_m})}{2\pi} 
\left(\frac{T_m}{T_{1bottom}}+\frac{T_m}{T_{2bottom}}\right) 
\sqrt{1-e_m^2} ,   
\label{a1bottom2}
\end{eqnarray}
where we used Eqs. ($\ref{r1}$) and ($\ref{r2}$). 

Determination of the semimajor axis is sensitive to 
measurement errors in the widths of the hills.  
This statement can be proven by using 
Eq. ($\ref{a1bottom2}$). 
For simplicity, we assume 
$\Delta_p \sim \Delta_m \sim \Delta$ and  
$T_{1bottom} \sim T_{2bottom} \sim T_{bottom}$, 
so that Eq. ($\ref{a1bottom2}$) can be reduced to 
\begin{equation}
a_{pm} \sim 
\frac{2}{\pi} R_s \sqrt{\Delta} 
\frac{T_m}{T_{bottom}} \sqrt{1-e_m^2} . 
\label{asim}
\end{equation}
The logarithmic derivative of this becomes 
\begin{equation}
\frac{da_{pm}}{a_{pm}} \sim 
\frac{dR_s}{R_s} 
+ \frac{d\Delta}{2\Delta} 
+ \frac{dT_m}{T_m} 
- \frac{dT_{bottom}}{T_{bottom}} 
- \frac{e_mde_m}{(1-e_m^2)} .  
\label{daa}
\end{equation}
We focus on $dT_m/T_m$ and $dT_{bottom}/T_{bottom}$, 
because we can expect much more accurate measurements 
for $R_s$ and $\Delta$, and the last term involving $e_m$ 
may be relatively small ($e_m<1$ and $de_m<1$). 
Time resolution in observations seems $dT_m \sim dT_{bottom}$, 
while $T_m \gg T_{bottom}$. 
Therefore, $dT_m/T_m \ll dT_{bottom}/T_{bottom}$, 
which means that accurate measurements of $T_{bottom}$ 
are crucial for the determination of $a_{pm}$. 

\begin{figure}
    \FigureFile(120mm,100mm){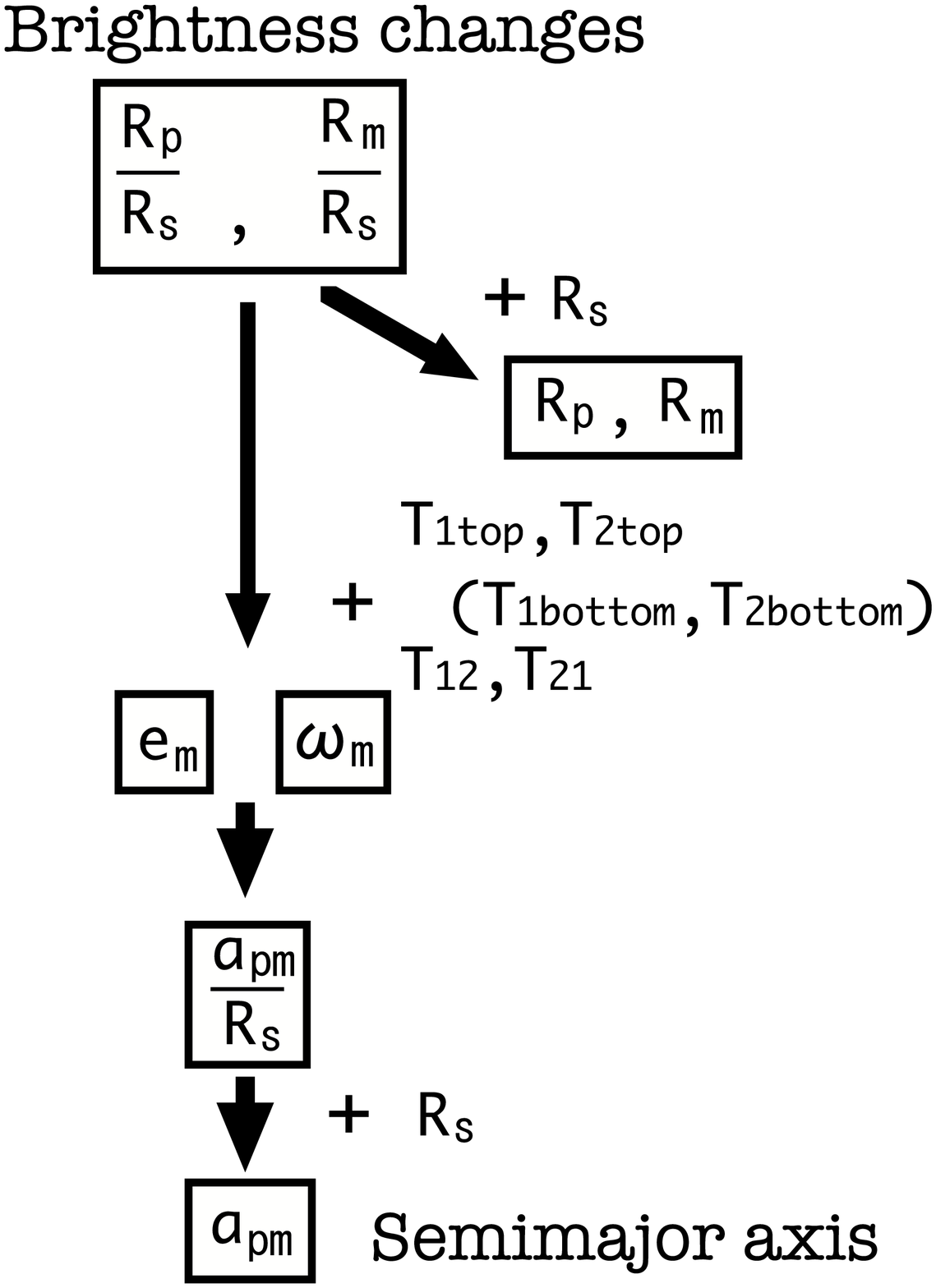}
\caption{ 
Flow chart of parameter determinations.  
Starting from measurements of brightness changes, 
the semimajor axis $a_{pm}$ can be finally determined 
for a fast case. 
}
\label{flow-chart}
\end{figure}

Figure $\ref{flow-chart}$ shows a flow chart of 
the parameter determinations that are discussed above.
The above formulation for parameter determinations 
actually recovers the correct values in Figs. 
$\ref{lightcurve-2}$-$\ref{lightcurve-4}$. 
In the numerical examples, the original parameters are retrieved 
within twenty percents. 

\begin{figure}
    \FigureFile(120mm,90mm){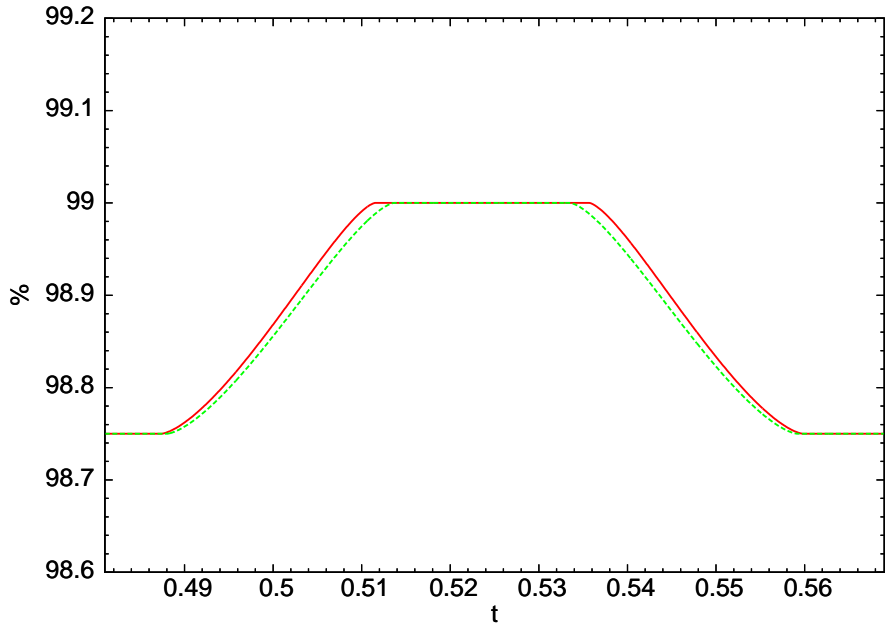}
\caption{
Difference in light curves due to orbital inclinations of a `moon'. 
All parameters but for the inclination angle are 
the same as those in Fig. $\ref{lightcurve-2}$. 
The solid (red) curve denotes $I_m = 90$ degree case, 
while the dashed (green) one means a case of $I_m = 88$ deg. 
Because of the orbital inclination, 
the duration of a mutual transit by the satellite is shortened. 
}
\label{lightcurve-inclination}
\end{figure}

\begin{figure}
    \FigureFile(120mm,100mm){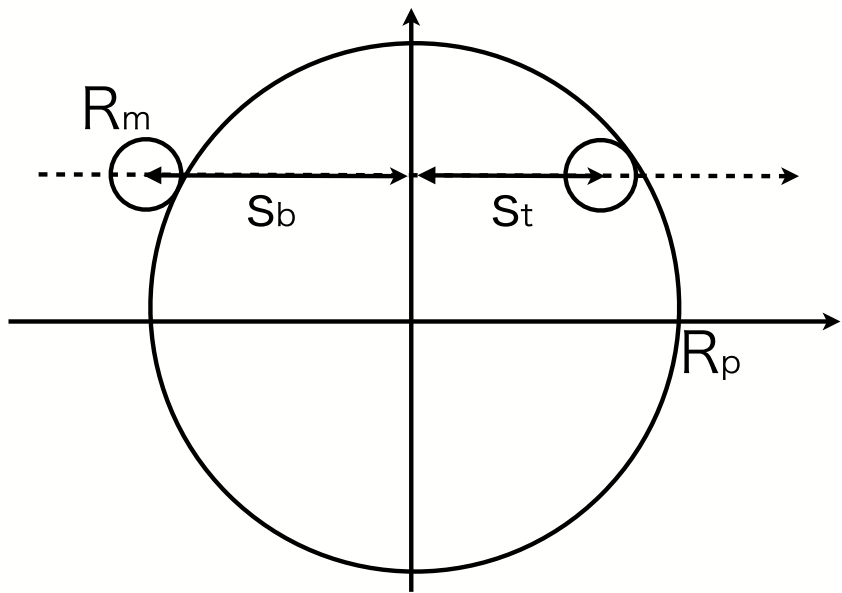}
\caption{ 
Definition of $s_b$ and $s_t$: 
The total transit duration $T_{bottom}$ is given by $s_b$, 
whereas the `flat part' of the hill $T_{top}$ is done by $s_t$. 
}
\label{definition}
\end{figure}

\noindent
{\bf Eccentric Orbit and Orbital Inclination:}\\
Figure $\ref{lightcurve-inclination}$ shows a difference between light curves 
for the edge-on ($I_m = 90$ deg.) and 
an inclination case ($I_m = 88$ deg.). 
Let $z$-axis denote the axis normal to the $x$-axis 
on the celestial sphere. 
We define the distance of a `moon' from the $z$-axis 
at the initial time of the mutual transit as 
\begin{equation}
s_b = \sqrt{(R_p+R_m)^2 
- \left(\frac{a_{pm}(1-e_m^2)}{1+e_m\cos f}\right)^2 \cos^2 I_m} , 
\label{sb}
\end{equation}
where the subscript $b$ means that the quantity is related 
with $T_{1 bottom}$ and $T_{2 bottom}$ as shown below 
(See also Fig. $\ref{definition}$). 
Similarly, when the `flat part' of the spike in light curves starts 
(or ends), we define the distance of a moon from the $z$-axis 
at this epoch as 
\begin{equation}
s_t = \sqrt{(R_p-R_m)^2 
- \left(\frac{a_{pm}(1-e_m^2)}{1+e_m\cos f}\right)^2 \cos^2 I_m} , 
\label{st}
\end{equation}
where the subscript $t$ means that the quantity is related 
with $T_{1 top}$ and $T_{2 top}$ as shown below. 

Therefore, we obtain the duration $T_{top}$ as 
\begin{equation}
T_{top} = \frac{2s_t}{V_f} ,  
\end{equation}
where $V_f$ is given by Eq. ($\ref{Vf}$). 
For the primary transit ($f=\pi/2 - \omega_m$), we thus obtain 
\begin{equation}
T_{1 top} = \frac{2s_{t1} \sqrt{1-e_m^2}}{a_{pm}n_m (1+e_m\sin\omega_m)} , 
\label{T1top-I}
\end{equation}
whereas for the secondary ($f=3\pi/2 - \omega_m$), we have 
\begin{equation}
T_{2 top} = \frac{2s_{t2} \sqrt{1-e_m^2}}{a_{pm}n_m (1-e_m\sin\omega_m)} . 
\label{T2top-I}
\end{equation}
Here, we define $s_{t1}$ and $s_{t2}$ as 
\begin{eqnarray}
s_{t1}&=&
\sqrt{(R_p-R_m)^2 
- \left(\frac{a_{pm}(1-e_m^2)}{1+e_m\sin\omega_m}\right)^2 \cos^2 I_m} , 
\label{st1}
\\
s_{t2}&=&
\sqrt{(R_p-R_m)^2 
- \left(\frac{a_{pm}(1-e_m^2)}{1-e_m\sin\omega_m}\right)^2 \cos^2 I_m} . 
\label{st2}
\end{eqnarray}
In the similar manner, we obtain the width of the 
spikes at the bottom as 
\begin{equation}
T_{bottom} = \frac{2s_b}{V_f} .  
\end{equation} 
For the primary transit ($f=\pi/2 - \omega_m$), we thus obtain 
\begin{equation}
T_{1 bottom} = \frac{2s_{b1} \sqrt{1-e_m^2}}{a_{pm}n_m (1+e_m\sin\omega_m)} , 
\label{T1bottom-I}
\end{equation}
whereas for the secondary ($f=3\pi/2 - \omega_m$), we have 
\begin{equation}
T_{2 bottom} = \frac{2s_{b2} \sqrt{1-e_m^2}}{a_{pm}n_m (1-e_m\sin\omega_m)} , 
\label{T2bottom-I}
\end{equation}
where we define $s_{b1}$ and $s_{b2}$ as 
\begin{eqnarray}
s_{b1}&=&
\sqrt{(R_p+R_m)^2 
- \left(\frac{a_{pm}(1-e_m^2)}{1+e_m\sin\omega_m}\right)^2 \cos^2 I_m} , 
\label{sb1}
\\
s_{b2}&=&
\sqrt{(R_p+R_m)^2 
- \left(\frac{a_{pm}(1-e_m^2)}{1-e_m\sin\omega_m}\right)^2 \cos^2 I_m} .
\label{sb2}
\end{eqnarray}

Because of the orbital inclination, we have to consider 
$T_{top}$ and $T_{bottom}$, separately. 
We define the ratios as 
\begin{equation}
T_{r top} \equiv \frac{T_{2 top}}{T_{1 top}} , 
\end{equation}
and 
\begin{equation}
T_{r bottom} \equiv \frac{T_{2 bottom}}{T_{1 bottom}} . 
\end{equation}
Substitutions of Eqs. ($\ref{T1top-I}$),  ($\ref{T2top-I}$), 
($\ref{T1bottom-I}$), ($\ref{T2bottom-I}$) into these ratios 
lead to 
\begin{eqnarray}
T_{r top}&=& \frac{s_{t2}}{s_{t1}} \frac{1+e_m\sin\omega_m}{1-e_m\sin\omega_m} ,
\label{Trtop-I}
\\
T_{r bottom}&=& \frac{s_{b2}}{s_{b1}} \frac{1+e_m\sin\omega_m}{1-e_m\sin\omega_m} .
\label{Trbottom-I}
\end{eqnarray}

For the edge-on case ($I_m = 90$ deg.), 
we obtain $s_{t2}/s_{t1} = s_{b2}/s_{b1} = 1$. 
Then, we have $T_{r top} = T_{r bottom}$. 
For a general case ($I_m \neq 90$ deg.), on the other hand, 
we find $T_{r top} \neq T_{r bottom}$. 
We thus expect that a ratio between them will 
give us the information about the orbital inclination. 
The ratio is 
\begin{equation}
\frac{T_{r top}}{T_{r bottom}} 
= \frac{s_{t2}}{s_{t1}} \frac{s_{b1}}{s_{b2}} .
\label{Tratio-I}
\end{equation}
The L.H.S. can be measured by observations. 

There are four unknown quantities $a_{pm}$, $e_m$, $I_m$ and $\omega_m$. 
We have four equations of 
($\ref{Trtop-I}$), ($\ref{Trbottom-I}$), ($\ref{Tratio-I}$) and 
the last one that can be chosen out of ($\ref{T1top-I}$), ($\ref{T2top-I}$), 
($\ref{T1bottom-I}$) and ($\ref{T2bottom-I}$). 
Therefore, one can determine the quantities $a_{pm}$, $e_m$, $I_m$ and
$\omega_m$ by using these equations for observations. 
For practical observations, data fittings at the slope 
of light curves are used, instead of transit durations, 
for determinations of the orbital inclination angle 
(Charbonneau et al. 2000). 
Nevertheless, an analytic solution is necessarily worthwhile 
to understand the properties of a given physical system, 
even when numerical fits are in practice the best way to 
determine the system parameters. 

\begin{figure}
    \FigureFile(120mm,90mm){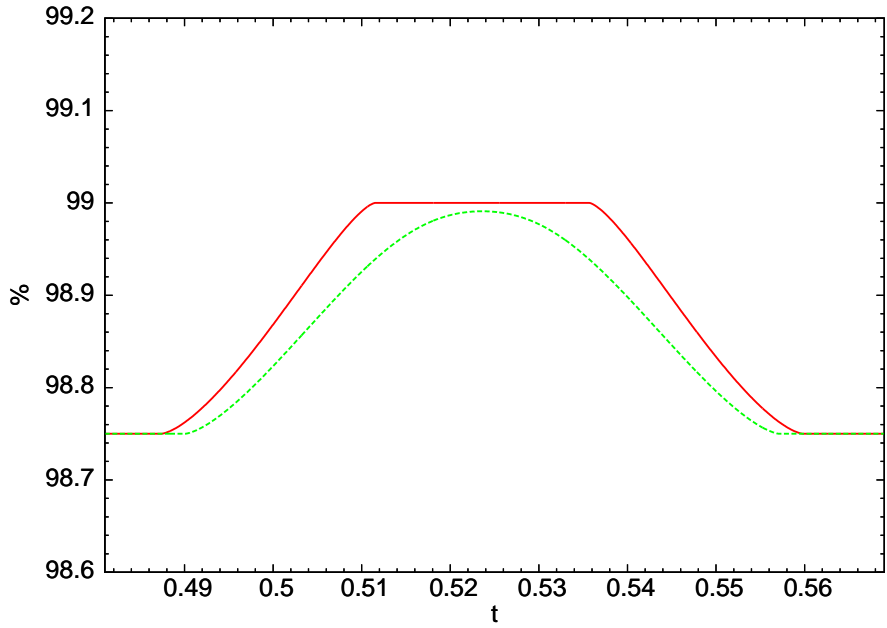}
\caption{
Light curve by a partial mutual transit of a satellite. 
All parameters but for the inclination angle are 
the same as those in Fig. $\ref{lightcurve-2}$. 
The solid (red) curve denotes $I_m = 90$ degree case, 
while the dashed (green) one means a case of $I_m = 86$ degree 
for a partial transit. 
Such a partial transit case produces 
a `U'-shaped hill in light curves. 
}
\label{lightcurve-partial}
\end{figure}

A partial transit occurs if the apparent impact parameter of 
the moon is in $(R_p-R_m, R_p+R_m)$. 
For the primary transit ($f=\pi/2 - \omega_m$), 
it occurs if the orbital inclination angle satisfies 
\begin{equation}
\frac{R_p-R_m}{a_{pm}} \frac{1+e_m\sin\omega_m}{1-e_m^2} 
< \cos I_m 
< \frac{R_p+R_m}{a_{pm}} \frac{1+e_m\sin\omega_m}{1-e_m^2} .
\end{equation}
For the secondary one ($f=3\pi/2 - \omega_m$), 
the condition of a partial transit becomes 
\begin{equation}
\frac{R_p-R_m}{a_{pm}} \frac{1-e_m\sin\omega_m}{1-e_m^2} 
< \cos I_m 
< \frac{R_p+R_m}{a_{pm}} \frac{1-e_m\sin\omega_m}{1-e_m^2} .
\end{equation}
Such a partial transit by a moon orbiting a host planet  
produces a `U'-shaped spike in light curves 
(See Fig. $\ref{lightcurve-partial}$).

\subsection{Timescales and brightness changes} 
We have presented a formalism for parameter determinations. 
Before closing this section, let us make brief comments 
on typical timescales and amplitudes in the brightness changes. 

The timescale of a brightness change due to a giant planet 
is about 
\begin{equation}
\frac{R_p}{a_{pm}n_m} 
\sim 
5 \times 10^3 
\left(
\frac{R_p}{5 \times 10^4 \mbox{km}}\frac{10 \mbox{km/s}}{a_{pm}n_m} 
\right) 
\mbox{sec.} 
\end{equation}
Therefore, {\it detections} of such fluctuations  
due to mutual transits of extrasolar planet-moon systems 
require frequent observations, 
say every hour. 
Furthermore, higher 
frequency (e.g., every ten minutes) 
is necessary for parameter {\it estimations} of the system. 

Let us mention a connection of the present result 
with space telescopes in operation. 
Decrease in apparent luminosity due to the secondary planet 
is $O(R_m^2/R_s^2)$. 
Besides the time resolution (or observation frequency) and 
mission lifetimes, 
detection limits by {\it COROT} 
with the achieved accuracy 
of photometric measurements (700 ppm in one hour) 
could put $R_m/R_s \sim 2 \times 10^{-2}$. 
The nominal integration time is 32 sec. but 
co-added over 8.5 min. except for 1000 selected targets 
for which the nominal sampling is preserved. 
By the {\it Kepler} mission 
with expected 
20 ppm differential sensitivity 
for solar-like stars with $m_V=12$, 
the lower limit will be reduced to $R_m/R_s \sim 4 \times 10^{-3}$. 
An analogy of the Earth-Moon 
($R_m/R_s \sim 2.5 \times 10^{-3}$, $W \sim 0.03$)
and 
Jupiter-Ganymede ($R_m/R_s \sim 4 \times 10^{-3}$, $W \sim 0.8$) 
will be marginally detectable. 
Observations both with high frequency 
(at least during the time of transits) 
and with good photometric sensitivity are desired 
for future detections of mutual transits.

\subsection{Dynamical limit and constraints on W} 
For Roche limit, we have 
\begin{equation}
a_{pm} < \beta R_H , 
\end{equation}
where $\beta$ denotes a numerical coefficient $0 < \beta <1$ 
and $R_H$ is Hill radius 
(See Domingos et al. 2006 for more detailed stability 
arguments by numerical computations). 
For simplicity, we assume $M_s \gg M_p \gg M_m$.  
Then, we have $a_{pm} \approx a_m$ and 
$R_H$ is approximated as 
\begin{equation}
R_H = \left( \frac{M_p}{3M_s} \right)^{1/3} d_p ,
\end{equation}
where $M_s$ denotes a host star mass and 
$d_p$ denotes the orbital radius of a planet 
orbiting the star. 
Kepler's third law gives $a_{pm}n_m$ and 
$v_{CM}$ as 
\begin{eqnarray}
a_{pm}n_m &=& \sqrt{\frac{GM_p}{a_m}} ,
\\
v_{CM} &=& \sqrt{\frac{GM_s}{d_p}} .
\end{eqnarray}
Combining these relations, therefore, we find 
\begin{equation}
W > 3^{1/6} \beta^{-1/2} \left( \frac{M_p}{M_s} \right)^{1/3} . 
\label{W>}
\end{equation}
If one assumes $M_p \sim M_J$ (Jupiter mass), 
we obtain $W > 10^{-1}$. 
This lower bound is less severe. 
On the other hand, 
there is a stringent constraint that the moon's closest approach 
$r_{min}$ 
cannot be within the planetary radius. 
We thus have 
\begin{equation}
a_m > \frac{r_{min}}{1-e_m} . 
\end{equation} 
This leads to 
\begin{equation}
W < \sqrt{ \frac{M_p}{M_s} \frac{d_p}{r_{min}} (1-e_m)} . 
\end{equation}
If one assumes the jovian mass and radius ($R_J$), 
this is rewritten as 
\begin{equation}
W < 1.4 
\left( \frac{d_p}{1 \mbox{AU}} \right)^{1/2} 
\left(\frac{M_{\odot}}{M_s}\right)^{1/2} 
\left(\frac{M_J}{M_p}\right)^{1/2}
\left( \frac{R_J}{r_{min}} \right)^{1/2}
(1-e_m)^{1/2} . 
\label{W<}
\end{equation}
For $W=6$ and $e_m=0.3$, we obtain $d_p > 25$ AU. 
Namely, a planet with a long orbital period $T_p > 125$ years 
is required. 
Kepler mission for several years is unlikely to 
see a transit by such a long period planet. 

Next, we consider a constraint that a mutual transit can occur. 
The transit duration for a planet in circular orbit is 
\begin{equation}
D \sim \frac{T_p R_s}{\pi d_p} , 
\end{equation}
where we assume the maximum duration by taking 
the vanishing impact parameter 
(See Seager and Mall{\'e}n-Ornelas for a more accurate form). 
As this limiting case for the Roche limit, 
we obtain the fastest case of a `moon' 
as 
\begin{equation}
a_m \sim \beta \left( \frac{M_p}{3 M_s} \right)^{1/3} d_p . 
\end{equation}
Using the Kepler's third law, this leads to 
\begin{equation}
T_m \sim \left(\frac{\beta^3}{3}\right)^{1/2} T_p . 
\end{equation}
For our fast case, we require $T_m < D$, 
which gives a bound on $\beta$ as 
\begin{equation}
\beta < 3^{1/3} \left( \frac{R_s}{\pi d_p}  \right)^{2/3} . 
\end{equation}
We substitute this into $\beta$ of Eq. ($\ref{W>}$) 
so that we can obtain 
\begin{eqnarray}
W&>&\left( \frac{\pi d_p}{R_s} \right)^{1/3}
\left( \frac{M_p}{M_s} \right)^{1/3} 
\nonumber\\
&\sim& 
1.5 
\left( \frac{d_p}{d_J} \right)^{1/3}
\left( \frac{R_{\odot}}{R_s} \right)^{1/3} 
\left( \frac{M_p}{M_J} \right)^{1/3}
\left( \frac{M_{\odot}}{M_s} \right)^{1/3}  
\nonumber\\
&\sim& 
0.9
\left( \frac{T_p}{1 \mbox{year}} \right)^{2/9}
\left( \frac{R_{\odot}}{R_s} \right)^{1/3} 
\left( \frac{M_p}{M_J} \right)^{1/3}
\left( \frac{M_{\odot}}{M_s} \right)^{1/3} , 
\label{W>2}
\end{eqnarray}
where $d_J$ denotes the mean orbital radius of the Jupiter. 

\begin{figure}
    \FigureFile(120mm,90mm){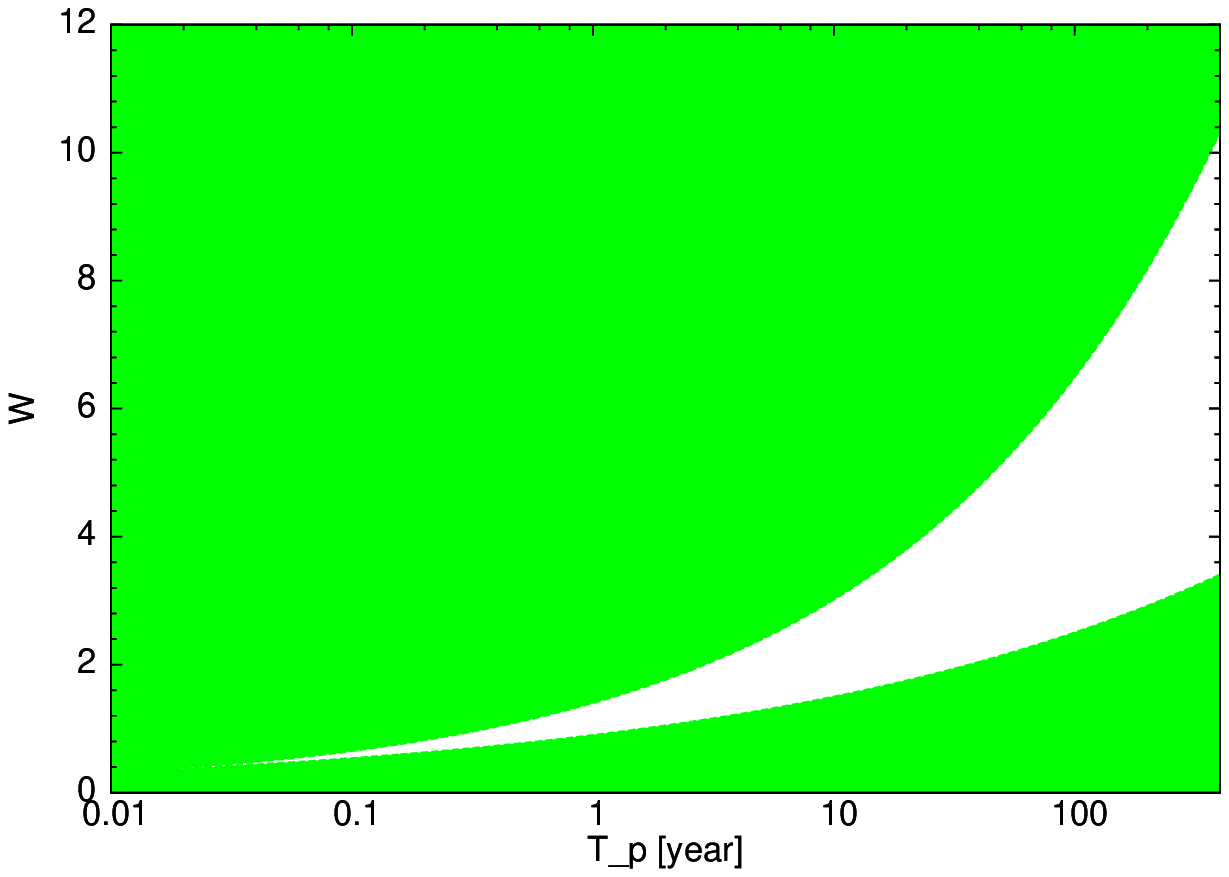}
\caption{
Possible bound on $W$ and $T_p$. 
Here, we assume $M_s = M_{\odot}$, $R_s = R_{\odot}$, $M_p = M_J$, 
$r_{min} = R_J$ and $e_m = 0$. 
The shaded (green) region denotes prohibited regions 
of the parameters that are constrained by 
Eqs. ($\ref{W>2}$) and ($\ref{W<2}$). 
}
\label{band}
\end{figure}

On the other hand, the dynamical arguments put 
an upper bound by Eq. ($\ref{W<}$). 
This is rewritten as
\begin{equation}
W < 1.4 
\left( \frac{T_p}{1 \mbox{year}} \right)^{1/3}
\left(\frac{M_{\odot}}{M_s}\right)^{1/2} 
\left(\frac{M_J}{M_p}\right)^{1/2}
\left( \frac{R_J}{r_{min}} \right)^{1/2}
(1-e_m)^{1/2} . 
\label{W<2}
\end{equation}
Therefore, there exists a narrow band 
as shown by Fig. $\ref{band}$. 
Outside this range, the proposed method cannot work.

\section{Conclusion} 
We have shown that light curves by mutual transits of 
an extrasolar planet with a `moon' depend on 
the moon's orbital eccentricity, 
especially for small separation (fast) cases, 
in which  
occultation of one faint object by the other 
transiting a parent star causes an apparent increase 
in light curves 
and such characteristic fluctuations with the same height 
repeatedly appear. 
We have also presented a formulation for determining 
the parameters such as the orbital eccentricity, inclination, 
semimajor axis and the direction of the observer's line of sight. 
This will be useful for probing the nature of the transiting 
planet-moon system. 

When actual light curves are analyzed, we should incorporate 
(1) photometric corrections such as limb darkenings, 
and 
(2) perturbations as three (or more)-body gravitating interactions 
(e.g., Danby 1988, Murray and Dermott 2000).

We would like to thank the referee for invaluable comments 
on the manuscript. 
We would like to thank S. Ida, S. Inutsuka. E. Kokubo and Y. Suto 
for stimulating conversations and encouragements. 
This work was supported in part (H.A) 
by a Japanese Grant-in-Aid 
for Scientific Research from the Ministry of Education, 
No. 21540252.


\clearpage
\begin{table}
\caption{
List of quantities characterizing a system in this paper. 
}
\label{table1}
  \begin{center}
    \begin{tabular}{ll}
Symbol & Definition \\
\hline 
$T_m$ & Orbital period of an extrasolar `moon' around a planet\\
\hline
$n_m$ & Mean motion of a moon   
$(=2\pi/T_m)$ \\
\hline 
$a_m$ & Semimajor axis of a moon's orbit w.r.t. 
a planet-moon's center of mass \\
\hline 
$a_p$ & Semimajor axis of a planet's orbit w.r.t. 
a planet-moon's center of mass \\
\hline 
$a_{pm}$ & Semimajor axis of a moon's orbit around a planet 
$(=a_p+a_m)$ \\
\hline 
$a_{\perp}$ & Apparent separation of a planet-moon system\\
\hline 
$R_s$ & Radius of a host star \\
\hline 
$R_p$ & Radius of a planet \\
\hline 
$R_m$ & Radius of a moon \\
\hline 
$M_p$ & Mass of a planet\\
\hline 
$M_m$ & Mass of a moon\\
\hline 
$M_{tot}$ & $M_p + M_m$\\
\hline 
$x_{CM}$ & Transverse position of a planet-moon's center of mass\\
\hline 
$v_{CM}$ & Transverse velocity of a planet-moon's center of
mass\\
\hline 
$x_p$ & Transverse position of a planet \\
\hline 
$x_m$ & Transverse position of a moon \\
\hline 
$e$ & Orbital eccentricity of the moon\\
\hline 
$t_0$ & Time of periastron passage of the moon\\
\hline 
$\omega_m$ & Argument of pericenter of the moon\\
\hline 
$t_{CM}$ & Time when the planet-moon's center of mass 
passes across the star's center\\
\hline 
$\Delta_p$ & Decrease rate in apparent brightness 
due to the planet transit\\
\hline 
$\Delta_m$ & Decrease rate in apparent brightness 
due to the moon transit\\
\hline 
$T_{1top}$ & Time duration: width of a hill's top at the primary
transit in light curves\\
\hline 
$T_{1bottom}$ & Time duration: width of a hill's bottom 
at the primary transit in light curves\\ 
\hline 
$T_{2top}$ & Time duration: width of a hill's top 
at the secondary transit in light curves\\
\hline 
$T_{2bottom}$ & Time duration: width of a hill's bottom 
at the secondary transit in light curves\\ 
\hline 
$T_{12}$ & Time lag from the first transit to the secondary\\ 
\hline 
$T_{21}$ & Time lag from the secondary transit to the first\\
\hline 
$\delta T$ & $T_{12}-T_{21}$ \\

    \end{tabular}
  \end{center}
\end{table}

\clearpage

\clearpage

 \clearpage

\clearpage

\clearpage

\clearpage

\clearpage

\clearpage

\clearpage

\clearpage

\begin{thebibliography}{}
\bibitem{Aitken}
Aitken, R. G. 1964 {\it The Binary Stars} 
(NY: Dover)
\bibitem{Binnendijk}Binnendijk, L. 1960 
{\it Properties of Double Stars} 
(Philadelphia: University of Pennsylvania Press) 
\bibitem{CS}
Cabrera, J., \& Schneider, J.\ 2007, A\&A, 464, 1133 
\bibitem{CW}
Canup, R.~M., \& Ward, W.~R.\ 2006, Nature, 441, 834
\bibitem{Transit}
Charbonneau, D., Brown, T.~M., Latham, D.~W., \& Mayor, M., 2000, 
ApJ, 529, L45
\bibitem{Danby}
Danby, J. M. A., 1988, 
{\it Fundamentals of Celestial Mechanics} 
(VA, William-Bell) 
\bibitem{Deeg1998}
Deeg, H.~J., et al., 1998, A\&A, 338, 479
\bibitem{Deeg2000}
Deeg, H.~J., Doyle, L.~R., Kozhevnikov, V.~P., Blue, J.~E., 
Martin, E.~L., Schneider, J., 2000, A\&A, 358, L5 
\bibitem{Transit2}
Deming, D., Seager, S., Richardson, L.~J., \& Harrington, J., 2005, 
Nature, 434, 740 
\bibitem{DWY}
Domingos, R.~C., 
Winter, O.~C., \& Yokoyama, T.\ 2006, MNRAS, 373, 1227
\bibitem{Doyle2000}
Doyle, L.~R., et al., 2000, ApJ, 535, 338 
\bibitem{Gaudi} 
Gaudi, B. S., \& Winn, J.~N., 2007, ApJ, 655, 550 
\bibitem{JH}
Jewitt, D., \& Haghighipour, N.\ 2007, ARAA, 45, 261 
\bibitem{JS}
Jewitt, D., \& Sheppard, S.\ 2005, 
Space. Sci. Rev., 116, 441 
\bibitem{Kipping09a}
Kipping, D.~M.\ 2009a, MNRAS, 392, 181 
\bibitem{Kipping09b}
Kipping, D.~M.\ 2009b, arXiv:0904.2565  
\bibitem{Leger}
L{\'e}ger, A., et al.\ 2009, A\&A, 506, 287L
\bibitem{MD}    
Murray, C. D., Dermott, S. F.,\ 2000,  
{\it Solar system dynamics} 
(Cambridge, Cambridge U. Press)
\bibitem{Narita07}
Narita, N., et al.\ 2007, PASJ, 59, 763 
\bibitem{Narita08}
Narita, N., Sato, B., 
Ohshima, O., \& Winn, J.~N.\ 2008, PASJ, 60, L1 
\bibitem{Ohta} 
Ohta, Y., Taruya, A., \& Suto, Y. 2005, ApJ, 622, 1118 
\bibitem{Queloz}
Queloz, D., et al.\ 2009, A\&A, 506, 303
\bibitem{SS}
Sartoretti, P., \& Schneider, J.\ 1999, A\&AS, 134, 553 
\bibitem{SA}
Sato, M., \& Asada, H. 2009, PASJ, 61, L29 
\bibitem{SM}
Seager, S., \& Mall{\'e}n-Ornelas, G.\ 2003, ApJ, 585, 1038 
\bibitem{SSS}
Simon, A., Szatm{\'a}ry, K., \& Szab{\'o}, G.~M.\ 2007, A\&A, 470, 727 
\bibitem{SSDS}
Szab{\'o}, G.~M., Szatm{\'a}ry, K., Div{\'e}ki, Z., 
\& Simon, A.\ 2006, A\&A, 450, 395 
\bibitem{WKW}
Williams, D.~M., 
Kasting, J.~F., \& Wade, R.~A.\ 1997, \nat, 385, 234 
\bibitem{Transit3}
Winn, J.~N., et al.\ 2005, ApJ, 631, 1215  
\end{thebibliography}
\end{document}